# Engineering Large Anisotropic Magnetoresistance in La$_{0.7}$Sr$_{0.3}$MnO$_3$ Films at Room Temperature

*Paolo Perna\*, Davide Maccariello, Fernando Ajejas, Ruben Guerrero, Laurence Méchin, Stephane Flament, Jacobo Santamaria, Rodolfo Miranda, Julio Camarero*

Dr. D. Maccariello[†], F. Ajejas, Dr. R. Guerrero, and Dr. P. Perna*
IMDEA Nanociencia, Campus de Cantoblanco, 28049 Madrid, Spain
*Corresponding author. E-mail: paolo.perna@imdea.org
[†]Present address: Unité Mixte de Physique CNRS-Thales. 1 Av. Augustin Fresnel 91767 Palaiseau. France

Dr. L. Méchin and Prof. S. Flament
GREYC (UMR6072) CNRS-ENSICAEN & Université de Caen Normandie, 6 Bd. de Maréchal Juin, 14050 Caen, France

Prof. J. Santamaria[†]
GFMC, Departamento de Física de Materiales, Facultad de Física, Universidad Complutense de Madrid, Campus Moncloa, 28040 Madrid, Spain
GFMC Instituto de Magnetismo Aplicado, Universidad Complutense, Las Rozas 28230 Madrid.

[†]Present address: Unité Mixte de Physique CNRS-Thales. 1 Av. Augustin Fresnel 91767 Palaiseau. France

Prof. R. Miranda, Dr. J. Camarero
IMDEA Nanociencia, Campus de Cantoblanco, 28049 Madrid, Spain
D.F.M.C. and Instituto "Nicolás Cabrera", Universidad Autónoma de Madrid, Campus de Cantoblanco, 28049 Madrid, Spain
Condensed Matter Physics Center (IFIMAC), Universidad Autónoma de Madrid, Campus de Cantoblanco, 28049 Madrid, Spain

**Abstract**
**The magnetoresistance (MR) effect is widely employed in technologies that pervade our world from magnetic reading heads to sensors. Diverse contributions to MR, such as anisotropic, giant, tunnel, colossal, and spin-Hall, are revealed in materials depending on the specific system and measuring configuration. Half-metallic manganites hold promise for spintronic applications but the complexity of competing interactions has not permitted**

the understanding and control of their magnetotransport properties to enable the realization of their technological potential. Here we report on the ability to induce a dominant switchable magnetoresistance in $La_{0.7}Sr_{0.3}MnO_3$ epitaxial films, at room temperature (RT). By engineering an extrinsic magnetic anisotropy, we show a large enhancement of anisotropic magnetoresistance (AMR) which leads to, at RT, signal changes much larger than the other contributions such as the colossal magnetoresistance (CMR). The dominant extrinsic AMR exhibits large variation in the resistance in low field region, showing high sensitivity to applied low magnetic fields. These findings have a strong impact on the real applications of manganite based devices for the high-resolution low field magnetic sensors or spintronics.

## 1. Introduction

Perovskite half-metallic manganites are considered very promising materials for next generation spintronics because of their high spin-polarization (almost 100%) [1] and large magnetoresistance (MR) response. [2-4] The wide variety of ground states, and in most cases the common perovskite structure, exhibited by complex transition metal oxides allows their combination in highly perfect epitaxial heterostructures. Moreover, interesting device concepts have resulted from multilayer structures where half metallic manganites are combined with other (multi) ferroic layers. [5-7] In spite of that, the technological promise of manganite based devices has not been fulfilled [8,9] mostly due to the complexity of the physical scenarios governing the interplay between a wide variety of coupled interactions. [10] Harnessing the magnetotransport responses is essential for device design and operation, yet important questions remain on the physical origin of the low-field MR in manganites. [11,12]

In manganites, transport is coupled to magnetism by a double exchange mechanism [13] and by the spin polarized nature of conduction electrons. Colossal magnetoresistance (CMR) is very large close to the metal-to-insulating transition (MIT). [2,12] Typically it overshadows other magnetoresistance contributions such as anisotropic magnetoresistance (AMR) due to spin-orbit (SO) coupling and largely the small Lorentz magnetoresistance (LMR), [14] governed by the electronic structure (which is several orders of magnitude smaller). In addition, due to the high spin polarization of the conduction band, spin-dependent scattering at grain-boundaries, [15, 16,17] domain-walls [18] and other magnetic inhomogeneities, [19] can be significant. Several types of MR concur in the measurements, and generally it is difficult to get a clear picture of its dependence on magnetization and/or current direction. Disentangling the origin of MR in manganites requires clear cut experiments to isolate the various contributions.

A problem for the application of the CMR in spintronics is its isotropic character at low magnetic fields. CMR depends monotonically on magnetic field independently of its direction and as a consequence it is non switchable. This is contrary to AMR which depends on the direction of magnetization with respect to current, and is thus intrinsically switchable and more amenable for spintronic applications. Signatures of switchable magnetoresistance at coercivity in magnetic field sweeps of manganites are typically very weak, and are due to the AMR which is mostly overshadowed by the CMR. [15,20] For this reason, AMR has remained poorly understood in manganites. However, important applications could be envisaged if we succeed to tailor the nano- and microstructure of the sample and to disentangle AMR from CMR.

In this Communication, we show that by inducing an extrinsic anisotropy (through the use of vicinal surfaces) a large AMR can be engineered in half-metallic $La_{0.7}Sr_{0.3}MnO_3$ (LSMO) films at room temperature (RT), as schematically shown in **Figure 1**. By combining simultaneous magnetization (vectorial

Kerr) and transport measurements we disentangle the different contributions to the total MR response and demonstrate that AMR is in fact the dominant contribution. As such, it can be tuned in sign and intensity by conveniently choosing the magnetization-current configuration.

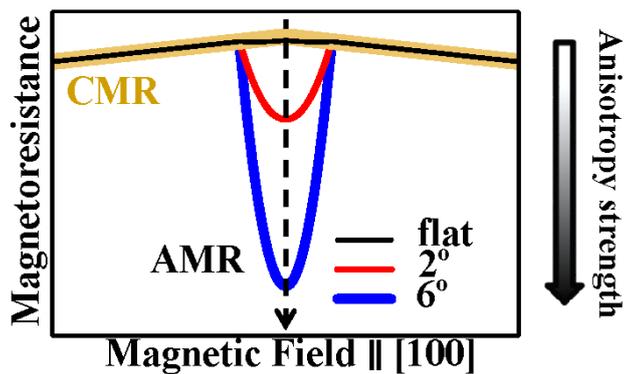

**Figure 1.** Schematic representation of engineering large AMR in LSMO films. By tailoring the magnetic anisotropy strength of LSMO (side arrow), through the use of substrates with progressively larger miscut angles [from 0º (flat) to 6º], despite the CMR contribution (yellow line) we are able to magnify over one order of magnitude the AMR signal.

AMR results from the SO interaction and its effect on the scattering between carriers and magnetic ions. Therefore, the sample resistivity depends on the angle between the sample magnetization and the applied current. In ferromagnetic 3d transition metal films, the AMR, which is computed as $[\rho_\parallel - \rho_\perp]/[(1/3)\rho_\parallel + (2/3)\rho_\perp]$, with $\rho_\parallel$ and $\rho_\perp$ being the in-plane resistivities for current parallel and perpendicular to the external magnetic field, largely dominates the overall MR response. [14] The two-current model based on SO interaction [21] that incorporates s-d electron scattering satisfactorily describes the AMR in metals, but it partially fails in reproducing the more complex scenario for manganites. In particular, while the AMR in metals is positive and monotonically dependent on temperature, [14] in manganites it is generally found to be negative [15] and non-monotonic [22] with temperature. In the latter compounds, AMR also has its origin in SO coupling ($H_{SO} = \lambda \, \mathbf{L} \cdot \mathbf{S}$), [15] although the complex interplay between electron, orbital, spin, and lattice degrees of freedom may affect the properties of the system near the phase transition. In fact, the orbital moment is completely quenched and for symmetry reasons, the matrix elements of the orbital momentum operator functioning on the $e_g$ states are zero (although they can be non-zero for $t_{2g}$ states as in the case of titanates). The SO coupling acts to second order in $\lambda/\delta E$, where $\lambda$ is the on-site SO interaction of Mn and $\delta E$ is the excitation (transfer) energy of $t_{2g}$ into $e_g$ levels. [15]

## 2. Results and discussion

In order to engineer a specific in-plane magnetic anisotropy we employed on purpose designed SrTiO$_3$ (STO) (001) vicinal surfaces, with miscut angle of 0º, 2º and 6º from the [001] towards [100] crystallographic direction, as substrates for the 30 nm thick LSMO epitaxial Pulsed Laser Deposition growth. These substrates are intentionally misoriented to a (near) low index surface, thus inducing terraces with edges along the [010] direction [see Supporting Information]. The resulting surface symmetry-breaking favors preferential anisotropy directions, defining a two-fold (uniaxial) in-plane magnetic anisotropy along the steps with anisotropy constant $K_U$. [23,24] Measurements of temperature dependent resistivity [$\rho(T)$], performed in a four-square contacts geometry and at zero-field showed low residual resistivity [$\rho(10\ K) \approx 10^{-6}\ \Omega m$] (which confirms a high crystal quality) [25], RT resistivity of about one order of magnitude larger, and Metal-to-Insulating transition (MIT) temperature above RT ($\approx 320\ K$) [see Supporting Information] in all samples, thus ensuring (ferromagnetic) metallic state at RT that is preferable for applications. [26]

The most efficient way to investigate the correlation between the magnetic and transport phenomena is to measure the field-driven magnetization and MR loops simultaneously. This experimental method has been previously exploited to investigate the MR response in ferromagnetic single layer, [27] ferromagnetic/antiferromagnetic bilayer [28] and spin-valve structures. [29] The sketches of the LSMO / STO (001) vicinal surface and of the combined vectorial-Kerr and MR measurement configuration are presented in **Figure 2**. In our vectorial-Kerr experiments, we measured simultaneously the in-plane parallel, $M_\parallel$, and transverse, $M_\perp$, magnetization components as function of the sample in-plane angular rotation angle ($\alpha_H$), and the MR for any field values and direction. In addition, the electrical current vector has been set either parallel or perpendicular to the anisotropy axis, i.e., $J_{010}$ and $J_{100}$ respectively.

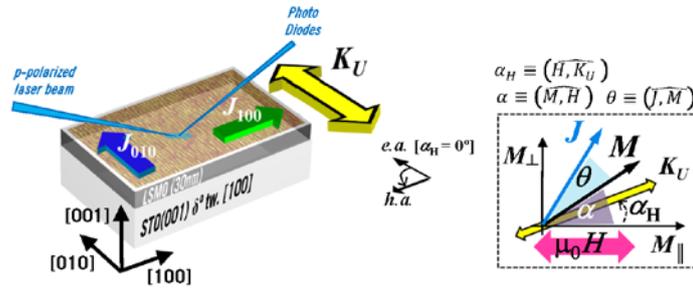

**Figure 2** Schematic representation of the LSMO film grown onto STO(001) substrate with miscut angle $\delta° = 0°$, $2°$ and $6°$ (from the [001] towards [100] crystallographic direction) and of the combined vectorial-Kerr and MR measurement configuration. In the left panel, an adapted atomic force microscopy (AFM) image (500x500 μm$^2$) of the LSMO vicinal surface that shows typical film grains elongated towards the direction of the substrate step-edges, i.e. along the [010]. The two-fold symmetry of the film morphology determines a defined two-fold (uniaxial) magnetic anisotropy with anisotropy constant $K_U$. The magnetization easy-axis (e.a.) is parallel to the [010] direction, whereas the hard-axis (h.a.) results perpendicular to it. In the side-box the illustration of the measurements configuration is sketched for clarity. It defines the angles between the magnetic field and the anisotropy direction $\alpha_H \equiv (\widehat{H, K_U})$, the magnetic field and the magnetization of the system $\alpha \equiv (\widehat{M, H})$ (i.e., magnetic torque), and injected current and magnetization $\theta \equiv (\widehat{J, M})$. Note that in our measurements, the magnetic field is kept fixed and the sample is rotated. The two orthogonal in-plane magnetization components are parallel ($M_\parallel$) and perpendicular ($M_\perp$) to the magnetic field $H$.

In **Figure 3** we have compared the RT MR responses (left panel) with the corresponding magnetization curves (right panels) of the LSMO films grown onto substrates with different miscut angles. We clearly observe that by increasing the miscut angle we get an enhancement of the MR signals, with magnetic field applied perpendicularly to the step edges (i.e., along

[100]). In particular, the largest MR variation (~0.28%) was obtained in the films deposited onto 6° miscut substrate (light-blue curve in panel a1), whereas the smallest MR (light-blue curve in panel c1) was measured in the flat films (0° miscut). The corresponding Kerr rotation field loops (proportional to the parallel-to-field magnetization component) [30] with magnetic field applied parallel (perpendicular) to the step-edges indicate that the films deposited on larger miscut angle substrate, present larger anisotropy field $H_K$, and larger MR. In contrast, the magnetic anisotropy of LSMO film deposited onto nominally flat substrate (i.e. 0° miscut) presents a weak magnetic anisotropy at RT, with a total MR one order of magnitude smaller (~0.03%). The small peaks at coercivity of MR curves in panel (c1) result from magnetization switching under the weak residual four-fold (biaxial) magneto-crystalline anisotropy [23b,31].

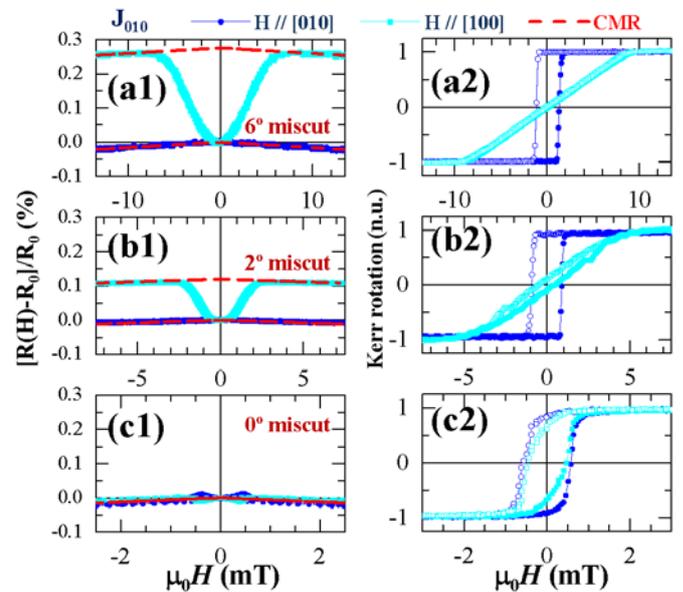

**Figure 3.** Left Panels: RT magnetoresistance responses of the LSMO films grown onto 6° (a1), 2° (b1) and 0° (c1) miscut STO (001) substrates with magnetic field applied along the [010] (blue curve) and [100] (cyan curve) direction. The MR curves are obtained by injecting the current along the anisotropy direction, i.e. along [010]. Red dashed curves are the fits of the CMR contribution. Right Panels: corresponding M-H loops with magnetic field applied along the [010] (blue curve) and [100] (cyan curve) direction. The anisotropy field $H_K$ is measured with magnetic field along the magnetization hard-axis, i.e. the [100] crystallographic direction. Note that the plot scales are adapted in order to appreciate the fine details. In particular, the MR of the 0º miscut sample (c1) is one order of magnitude smaller than the MR of 6º miscut sample (a1).

From the resistance changes loops in Figure 3 with field applied along the [010] (blue) and [100] (light-blue) of stepped LSMO (panels a1 and b1), it is possible to envisage two main regimes. At high magnetic fields, the film resistance varies linearly with H ($\propto -|\mu_0 H|$) independently to the direction of the magnetic field. At low fields, the two curves are completely different: for $\mu_0 H \parallel$ [010] [i.e., e.a.], R(H) shows the same linear behavior observed at high field, while for $\mu_0 H \perp$ [100] [i.e., h.a.] it presents a huge variation upon sweeping the field from positive to negative values, being minimum at zero field and maximum at high field ($\geq \mu_0 H_K$). The behavior at high magnetic fields, i.e. a linear drop of the resistivity as the magnetic field increases, is due to CMR effect. It does not depend on the direction of the external field and is maximum at the Curie temperature (above RT in our films). Note that a Lorentz magnetoresistance contribution to the measured MR can be discarded since it may produce a parabolic field-dependent resistance [14] that is not observed in our measurements. The behavior at low magnetic fields is dominated by AMR, which is due to the mixing of spin-↑ and spin-↓ states because of SO interaction. It depends on the relative orientation between the magnetization **M** and the injected electrical current **J**, and is generally described by a $\cos^2\theta$, [21,22] with $\theta (H) \equiv (\widehat{M,J})(H)$. The sign of the AMR in manganites, which is opposite to what found in 3d ferromagnetic metals, [27] derives from the $L_z S_z$ term of the SO interaction that leads to $e_g\uparrow$ states splitting. [15]

On the basis of aforementioned considerations, since **M** at the e.a. always lies parallel to the magnetic field [Figure 3 (a2,b2)], the corresponding R(H) [in Figure 3 (a1,b1)] should not show any AMR variation (**M** ∥ **J**, thus constant AMR). In clear contrast, at the h.a. **M** rotates during the field loop (M-H is fully reversible), and consequently the AMR contributes the most to R(H) changes. For applied fields larger than $\mu_0 H_K$, **M** is forced to be parallel to the field and, therefore, perpendicular to **J**. As the field decreases, the magnetization rotates and the angle $\theta$ changes continuously. At zero-field, **M** is oriented along the anisotropy axis, thus aligned to **J**. Note that for large magnetic field the CMR has the same slope for any field angle [red dashed line in Figure 3 (a1,b1,c1)]. This means that CMR is independent to the field direction, allowing the discrimination from the AMR contribution. It is worth remarking that in the case of flat film (i.e., 0º miscut in panel c1) the AMR is completely overshadowed by the CMR signal.

We compare the R(H) curves measured at RT by injecting the electrical current (and probing the voltage drop) in-plane either along the [010] (i.e., parallel to $K_U$) or along [010] (i.e., perpendicular to $K_U$), i.e., $J_{010}$ and $J_{100}$ respectively. This is reported in **Figure 4** in which we immediately see that: *i)* the sign of the resistance variation changes depending on the current direction; *ii)* the CMR contribution does not depend neither to the current nor the field direction.

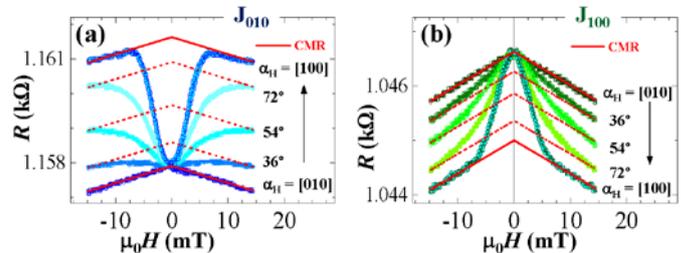

**Figure 4.** MR measurements at RT of LSMO film deposited onto 6º vicinal STO(001) for different magnetic field directions, from the [010] (i.e., e.a.) to the [100] (i.e., h.a.) crystallographic directions. Panel (a) corresponds to the case of current injected along the [010], $J_{010}$; panel (b) to the case of current injected along the [100], $J_{100}$. The red continuous lines are fits of the measured MR in the linear regions, which highlight the CMR contribution. Note that the CMR has the same slope for any field and current direction.

In order to gain further insight into the origin of the magnetotransport properties and to elucidate the role of the AMR in manganites, we have studied accurately the magnetization reversal pathways for any magnetic field values and directions [$\alpha_H \equiv (\widehat{H, K_U})$. $\alpha_H = 0º$ refers to the external field $\mu_0 H$ parallel to the anisotropy $K_U$ axis (i.e., [010] direction) (see sketch in Figure 2). The magnetization components, parallel ($M_{\parallel}$) and perpendicular ($M_{\perp}$) to the external field, are derived from vectorial-resolved magneto optic Kerr effect measurements.[30] **Figure 5** shows M-H and MR-H hysteresis loops (acquired simultaneously) at RT for selected directions of the magnetic field, from easy- to hard-axis direction of the LSMO film deposited onto 6º vicinal STO (001). The CMR contribution ($\propto -|\mu_0 H|$) has been extracted from the measured MR(H) loops, by fitting the R(H) curves in the linear region obtaining ≈ 0.04% at 20 mT. In order to isolate the AMR from the CMR, we have subtracted such linear contribution from the MR(H) curves acquired in the whole angular range (Figure 4).

From a simple inspection of the angular-dependent data, we notice that the MR-H curves change accordingly to the M-H, indicating their intimate correlation with the magnetic anisotropy of the system. We first focus on the field-driven magnetization behaviors in order to understand the influences of the magnetic symmetry on the magnetic properties of our system [Figure 5 (a,b)]. Then, we correlate the reversal mechanisms to the MR-H loops [Figure 5 (c,d)].

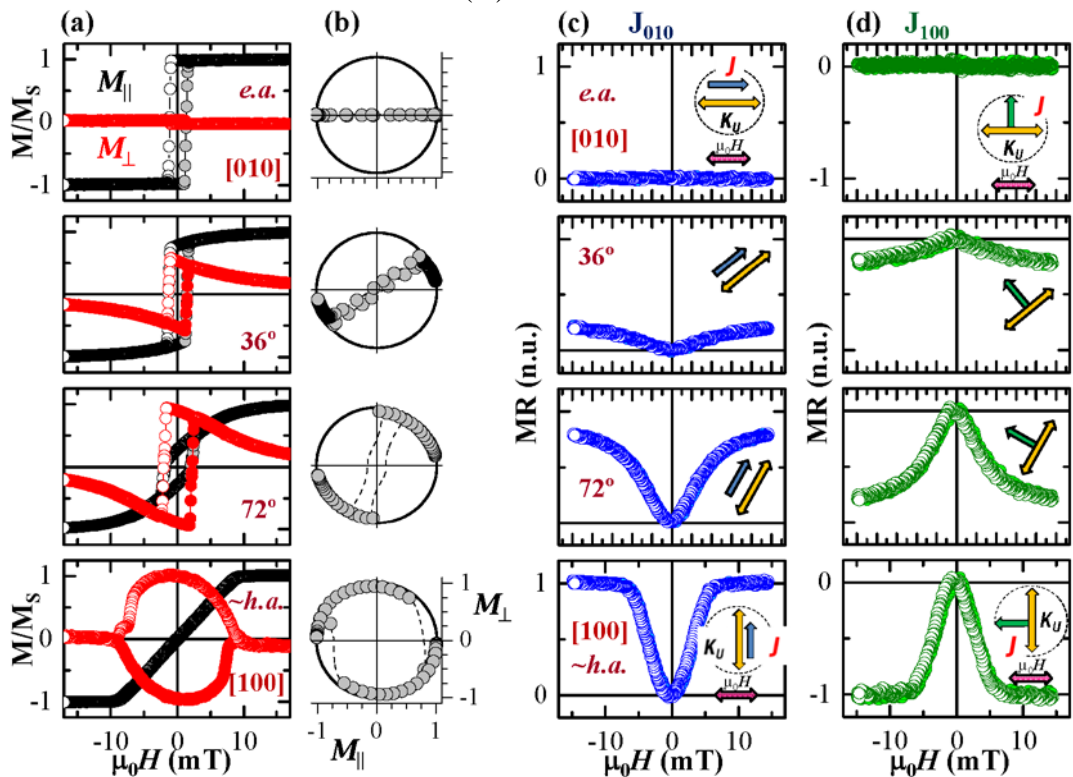

**Figure 5.** Simultaneous RT vectorial-Kerr and low-field MR curves of LSMO film deposited onto 6º vicinal STO(001), for selected direction of the magnetic field $\mu_0 H$ from e.a. ([010], $\alpha_H = 0º$) to h.a. ([100], $\alpha_H = 90º$). Panel (a) shows the $M_{\parallel}/M_S$-H and $M_{\perp}/M_S$-H loops. At e.a., $M_{\parallel}$ presents a squared loop with sharp transitions, meanwhile $M_{\perp}$ is negligible. Approaching to the h.a. smoother transitions in both magnetization components become progressively significant. At the h.a., $M_{\parallel}$ is fully reversible and $M_{\perp}$ describes a quasi-perfect circle. In (b) $M_{\perp}$ vs. $M_{\parallel}$ at the corresponding $\alpha_H$ are reported. These plots allow to clearly see that the sample magnetization always follows the anisotropy axis: at e.a. M can only lie parallel to H, whereas close to h.a. it rotates during the reversal. Panels (c) and (d) present MR= [R(H)−R$_0$]/R$_0$ curves for current injected along the [010] direction, i.e., $J_{010}$, and along the [100], i.e., $J_{100}$, respectively. MR curves have been corrected for the presence of CMR (see text). When the field is applied along $K_U$ the MR is constant in both cases, being minimum (maximum) for $J_{010}$ ($J_{100}$). In both configurations, the largest MR variation occur when the field is applied perpendicular to $K_U$ (i.e., h.a.) because $\theta$ varies smoothly from 0º to 180º. Note that positive and negative MR signals are obtained depending on the magnetization-current configuration.

With this aim we use two representations, standard M-H [Figure 5(a)] and polar $M_\perp$ vs. $M_\parallel$ [Figure 5(b)] plots to identify the preferential magnetization direction, critical fields, domain-wall angles and magnetization reversal processes. Figure 5(a) shows $M_\parallel$-H and $M_\perp$-H loops (normalized to the saturation magnetization $M_S$) for selected directions of the magnetic field $\alpha_H$. As mentioned above, exactly at the e.a. (i.e., $\mu_0H \parallel [010]$, $\alpha_H = 0°$) $M_\parallel$ presents a squared loop with sharp transitions, whereas $M_\perp$ does not vary at all. This is because the magnetization of the system switches from one direction to the other, always following the external field. This is a typical behavior of an e.a. region where the reversal is dominated by the nucleation and further propagation of magnetic domains oriented parallel to the external field. [23] When the field is misaligned with respect to $K_U$, smoother transitions in both magnetization components appear [e.g., at $\alpha_H = 36°$ and $72°$ in Figure 5(a)]. These signify that magnetization reversal processes become progressively more significant when approaching to the h.a. In other words, **M** tries to be aligned to the anisotropy direction, while it is parallel to the external field only if the latter is larger enough (at saturation). Close to the h.a., i.e. $\mu_0H \parallel [100]$ ($\alpha_H = 90°$), $M_\parallel$ becomes fully reversible (with no hysteresis) whereas $M_\perp$ describes a quasi-circular loop, meaning that magnetization rotation mechanisms are dominating the reversal. [24]

The $M_\perp$ vs. $M_\parallel$ polar-plots in Figure 5(b) (normalized to $M_S$) allow for the visualization of the in-plane trajectory of the magnetization vector during reversal. In this way, the specific mechanism of the magnetization reversal is easily elucidated. The data lying on the circle of unit radius [solid line in panel (b)], represent rotation processes. Every time the data are off this circle, magnetic domains are present. As the field is decreased from saturation, the magnetization vector rotates reversibly along the circle, except for e.a. The rotation continues for negative fields until a new irreversible process occurs (this is indicated by the deviation from the circle of the magnetization vector). Both departure and return points are close to the anisotropy axis, which mean 180° reversal. Hence, the sharp transitions correspond to reversal via nucleation of magnetic domains oriented along the anisotropy direction and further 180° domain-wall propagation. In contrast, at h.a., M describes a quasi-perfect circle proving that the reversal is governed by rotation mechanisms.

The corresponding MR curves are presented in Figure 5(c,d). Here, MR is defined as $[R(H)-R_0]/R_0$, where $R_0$ is the resistance at zero field. In panel (c), we show the case of $J_{010}$ injected along $K_U$ (i.e., along the [010] direction); in (d), the case of $J_{100}$ perpendicular to $K_U$ (i.e., along the [100]). These two configurations are sketched in Figure 2 for clarity. At first glance, we notice that for both current configurations, when the field is applied along the e.a. ($\mu_0H \parallel [010]$, $\alpha_H = 0°$) MR is constant in the field loop, but minimum for $J_{010}$ (because **M** and **J** are always parallel, as indicated in the side-sketch) and maximum for $J_{100}$ (because **M** and **J** are always perpendicular). For $\mu_0H \parallel [100]$ (i.e., h.a. direction), the MR-H loops show the largest variation (of the whole angular range) in both configurations, although with inverted sign. These behaviors are due to the smooth changes (from 0° to 180°) of $\theta$ (angle between **M** and **J**) during the field loop. Therefore, the angular range where magnetization rotation processes are more relevant (i.e., close to h.a. direction) correspond always to the largest MR changes. For the $J_{010}$ case we found MR=+0.28%, while for $J_{100}$ we obtained −0.16%.

In summary, the shape of the MR-H loop depends on the specific magnetization reversal pathway, and its value can be tuned from positive to negative by modifying the measuring conditions. For field-direction in which no **M** rotation occurs (i.e., at the e.a.), the resulting MR is constant, and is maximum or minimum depending on the direction of **J**. For field-direction in which the magnetization rotates (i.e., away from the e.a.), since the angle between the magnetization and current vectors changes, MR also varies. The largest MR variation is found therefore when the magnetic field is applied along the h.a. direction. In this configuration, in fact, the angle θ varies gradually from 0º to 180º during the field loop.

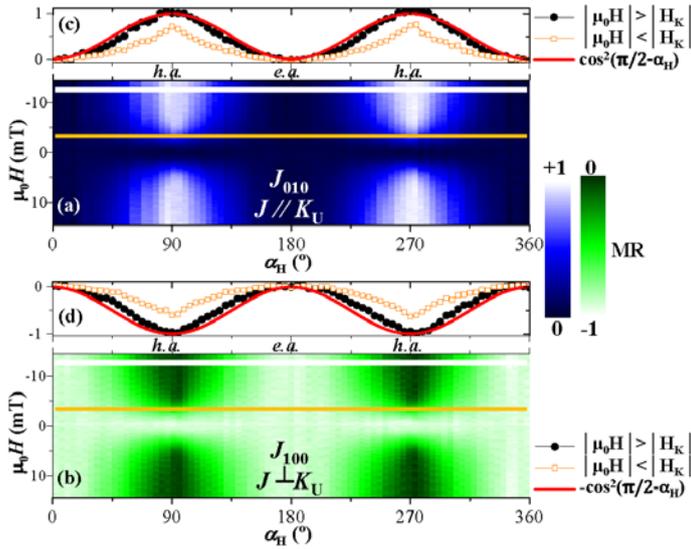

**Figure 6.** Two-dimensional map representations of the RT angular evolution of the forward branches of the MR curves of LSMO film deposited onto 6º vicinal STO(001). Panel (a) shows the case of current injected along the [010] direction ($J_{010}$), panel (b) the case of current injected along the [100] ($J_{100}$). This graphic representation allows the identification of two-fold (uniaxial) symmetry with 180º periodicity. By cutting the maps horizontally we get an angular evolution of MR at fixed magnetic field. By doing so, at field larger than the anisotropy field [$\mu_0 H$ = -13 mT, black-circles in insets (c,d)] we obtain a good agreement with $\cos^2 \alpha_H$, whereas for smaller field [$\mu_0 H$ = -3.7 mT, orange-squares in insets (c,d)] such a dependence is not valid.

By analyzing the anisotropy induced magnetotransport symmetry, i.e. the field direction dependence of the MR, we finally show that the low-field MR output in our system follows the $\cos^2 \theta$ law, i.e. it is due to AMR. The two-dimensional map representation of the angular MR evolution allows a clear picture of the MR dependence with field for both $J_{010}$ [**Figure 6(a)**] and $J_{100}$ [**Figure 6(b)**] current configurations. By cutting the map vertically, we get a MR-H loop at given field-direction, whereas the angular evolution of the low-field anisotropic MR at fixed values of the magnetic field is obtained by cutting the map horizontally. The latter is shown in the insets (c,d) for different field values. For $\mu_0 H$ >|$\mu_0 H_K$| = 9 mT, the curve resembles the $\cos^2 \alpha_H$ behavior [e.g., at $\mu_0 H$ = −13 mT, black-circles in the insets (c,d) of Figure 6], with inverted sign depending on the chosen magnetization-current configuration. At smaller field (e.g., at $\mu_0 H$ = −3.7 mT, orange-squares in insets) such a dependence is no longer satisfied, similarly to the case of 3d metals, [27] although the $\cos^2 \theta$ dependence is still valid.

## 3. Conclusions

In conclusion, we have exploited vicinal surfaces to engineer an extrinsic (uniaxial) magnetic anisotropy in LSMO which dominates magnetotransport at RT. By simultaneously measuring the magnetization and magnetoresistance hysteresis loops we have established the link between magnetization reversal pathways and anisotropic resistance changes. We have found that AMR dominates over CMR and any other spin dependent contribution due to grain-boundaries, domain-walls, inhomogeneities, etc. Apart from its sign (opposite to the metals case), the AMR in manganites behaves similarly to the AMR in metals, hence suggesting a similar underlying physics despite the complexity of the magnetic interactions in correlated oxides. The ability to engineer a switchable magnetoresistance in manganites at room temperature could open the way to new applications as high-resolution low field magnetic sensors in a future oxide electronics or spintronics.

## 4. Experimental Section

The LSMO thin films, with thickness of 30 nm, were deposited by pulsed laser deposition (PLD) from a stoichiometric target onto commercially available STO (001) substrates with miscut angle of 0º, 2º and 6º from the [001] towards [100] crystallographic direction. Details of the growth, structural, transport and morphological characterizations are reported in the Supporting Information. For the RT vectorial-Kerr experiments, we used p-polarized light (with 405 nm wavelength) focused on the sample surface and analyzed the two orthogonal components of the reflected light. This provides the (additional) simultaneous determination of the hysteresis loops of both in-plane parallel, $M_\parallel$, and transverse, $M_\perp$, magnetization components as function of the sample in-plane angular rotation angle ($\alpha_H$), keeping fixed the external magnetic field direction. Details on the experimental vectorial Kerr set-up can be found in Ref. [28]. The study of the magnetic anisotropy of the films is reported in the Supporting Information. Simultaneously to the vectorial-Kerr magnetization curves, and for any field values and direction, the MR was measured by using a lock-in amplifier in a four probe geometry. The electrical current vector has been set either parallel or perpendicular to the anisotropy axis, i.e., $J_{010}$ and $J_{100}$ respectively. We have used an ac current of about 2 mA with 39 kHz modulation. More details on the simultaneous MR and magnetization measurements can be found in Refs. [27,28,29]

**Supporting Information**
Supporting Information is available from the Wiley Online Library.

**Acknowledgements**
This work was supported in part by the Spanish MINECO through Projects No. MAT2012-39308, FIS2015-67287-P and PCIN-2015-111, and by the Comunidad de Madrid through Project NANOFRONTMAG CM. P.P. acknowledges support through the Marie Curie AMAROUT EU Programme and JCI-2011-09602. The authors acknowledge support from the CNRS through PICS2012 MRLSMO project. This project has received funding from the European Union's Horizon 2020 research and innovation programme under grant agreement No. 737116 (ByAXON).

**Author contributions**
P.P. and J.C. designed the experiment; D.M., F.A., J.C. and P.P. performed the measurements and analyzed the results, with the help of R.G.; L.M. and S.F. fabricated the samples; P.P. wrote the manuscript, with the help of D.M., J.C., L.M., S.F., J.S. and R.M. All authors discussed the results.